\documentclass[conference]{IEEEtran}
\IEEEoverridecommandlockouts

\usepackage{tikz}
\usepackage{pgfplots}

\usepackage[figuresright]{rotating}

\usepackage{amsmath}
\usepackage{amssymb}

\usepackage{color, colortbl}

\usepackage{algorithm,algpseudocode}
\usepackage{setspace}

\usepackage[dvipsnames]{pstricks}
\usepackage{pst-node}

\definecolor{table-gray}{gray}{0.5}
\definecolor{dark-gray}{gray}{0.3}

\newtheorem{theorem}{Theorem}

\newtheorem{lemma}{Lemma}

\def\qed{\hbox{${\vcenter{\vbox{			
   \hrule height 0.4pt\hbox{\vrule width 0.4pt height 6pt
   \kern5pt\vrule width 0.4pt}\hrule height 0.4pt}}}$}}

\title{Performance Comparisons of Self-stabilizing Algorithms for Maximal Independent Sets}

\author{
    \IEEEauthorblockN{Barton F. Cone\IEEEauthorrefmark{1}, Stephen T. Hedetniemi\IEEEauthorrefmark{2}, Lance C. Ingle\IEEEauthorrefmark{3}, Ken Kennedy\IEEEauthorrefmark{1}}
    \IEEEauthorblockA{\IEEEauthorrefmark{1}Fulfilld.io
    \\\{bart, ken\}@fulfilld.io}
    \IEEEauthorblockA{\IEEEauthorrefmark{2}School of Computing, Clemson University
    \\hedet@clemson.edu}
    \IEEEauthorblockA{\IEEEauthorrefmark{3}GTRI, Georgia Tech\\lanceingle@gtri.gatech.edu}
}

\begin{document}

\maketitle
\thispagestyle{empty}
\pagestyle{empty}

-----------------

\begin{abstract}

Sensor networks, such as ultra-wideband sensors for the smart warehouse, may need to run distributed algorithms for automatically determining a topological layout. In this paper, we present 5 different self-stabilizing algorithms (their central and distributed counterparts) for determining maximal independent sets. The performance of the algorithms, in terms of time complexity, simulation analysis, and size of maximal independent sets found are then compared.

\end{abstract}


\begin{section}{Introduction}\label{Introduction_Section}

Maximal independent sets (MIS) have several uses, one of which is in wireless networks \cite{AlzWan,wang}. Wireless and sensor networks are ideal testbeds for self-stabilizing algorithms, and this corresponds with work that Fulfilld.io is doing with ultra-wideband (UWB) sensors for the smart warehouse. The UWB sensors are positioned such that objects can be tracked with near realtime precision allowing a variety of algorithms to be performed (path planning, route optimization, bin optimization, ...).

In this paper, we compare a variety of self-stabilizing algorithms that have been designed to find independent sets, or maximal independent sets, of nodes in distributed systems.  The topology of a distributed system is modeled by an undirected graph $G = (V,E)$ of {\it order} $n = |V|$ of nodes, and {\it size} $m = |E|$ of edges, where nodes represent processes and edges represent communication links between processes.

The {\it open neighborhood} of a node $i \in V$ is the set $N(i) = \{ j | ij \in E\}$ of nodes that are adjacent to $i$; these nodes are called the {\it neighbors} of node $i$. The {\it closed neighborhood} of a node $i$ is the set $N[i] = N(i)  \cup \{i\}$. The {\it degree} of a node is $deg(i) = |N(i)|$, and the maximum degree of a node is denoted $\Delta(G)$.

The processes of a self-stabilizing algorithm consist of a set of {\it variables} associated with a node and a set of {\it rules}, each of which consists of a {\it guard} and a {\it statement}.  In the {\it distance-$1$ model}, a guard is a Boolean expression over the variables in the closed neighborhood $N[i]$ of a node $i$. If the Boolean expression evaluates to {\it true} at a node $i$, then node $i$ is said to be {\it eligible to make a move}, which consists of changing the values of the node's variables as specified by the associated statement.

The values of a node's variables at any point define its {\it local state}.  The local states of the $n$ nodes define the {\it global state} or {\it configuration} of the system.   An algorithm is called {\it self-stabilizing} if regardless of a current, initial configuration, the system is guaranteed to reach a {\it legitimate} or {\it desired} global state in a finite amount of time without any external intervention, and to remain in a legitimate state \cite{vt07}.

A set $S \subseteq V$ of nodes is called {\it independent} if no two nodes in $S$ are adjacent.  In the graph theory literature, $i(G)$ denotes the minimum cardinality of a maximal independent set, and $\beta_0(G)$ denotes the maximum cardinality of an independent set in $G$.

A set $S$ of nodes is called a {\it dominating set} if every node in $V \setminus S$ is adjacent to at least one node in $S$.  The {\it domination number} of a graph $G$ equals the minimum cardinality of a dominating set in $G$, while the {\it upper domination number} equals the maximum cardinality of a minimal dominating set in $G$.  The following inequality chain is well-known in domination theory \cite{hedet}.

\begin{equation}
  {\gamma \leq i \leq \beta_0 \leq \Gamma}
\end{equation}

We focus on comparing the performance of several self-stabilizing graph algorithms that have been published for finding maximal independent sets. In particular, we will simulate each algorithm in order to see which algorithms find either smaller or larger maximal independent sets.  We are also interested in the general question of whether it is possible to modify these algorithms so that they find maximal independent sets whose cardinalities are better approximations to either $i(G)$ or $\beta_0(G)$.  It should also be pointed out that maximal independent sets are also minimal dominating sets.  Therefore, these algorithms can be used to find minimal dominating sets whose cardinalities may approximate either $\gamma(G)$ or $\Gamma(G)$. 

\end{section} 


\begin{section}{Central Algorithms}\label{CS_Section}

In this section, we present five self-stabilizing algorithms for finding independent sets that utilize the distance-1 model with an unfair central scheduler, whereby only one eligible node ${x}$ is permuted to move at a time.  All of the algorithms are silent, in that after stabilization, no node will make another move. Since Algorithm C2 is a modification of Algorithm C1, and Algorithms C3 and C4 are further modifications of Algorithm C2, proofs of correctness and complexity are provided where relevant. Complexity results are stated for the other algorithms; the reader is referred to the original papers for their proofs.

\begin{subsection}{Central Scheduler Algorithm 1}\label{CS_1_Subsection}

Algorithm C1, due to Shukla, Rosenkrantz and Ravi \cite{Shukla}, was the first self-stabilizing algorithm for finding a maximal independent set. Each node $i$ maintains only one Boolean variable $x(i)$.  When the algorithm stabilizes, the set of nodes $X = \{ i$ $ |  x(i) = 1\}$ defines a maximal independent set.

\begin{algorithm}[htbp]
 \begin{algorithmic}
   \State \textbf{RIn: if} x(i) = 0 ${\wedge}$ (${\nexists}$j ${\in}$ N(i))(x(j) = 1)
   \State ${\;}$ \textbf{then} x(i) = 1

   \State \textbf{ROut: if} x(i) = 1 ${\wedge}$ (${\exists}$j ${\in}$ N(i))(x(j) = 1)
   \State ${\;}$ \textbf{then} x(i) = 0
 \end{algorithmic}
 \caption{C1}
 \label{alg:c1}
\end{algorithm}

\begin{theorem} Algorithm C1 finds a maximal independent set in at most 2n moves.
\label{lem:c1theorem} 
\end{theorem}
\noindent
\textbf{Proof:} See proof of Algorithm 3.1 in \cite{shsh03}.
\qed

\end{subsection}

\begin{subsection}{Central Scheduler Algorithm 2}\label{CS_2_Subsection}

Algorithm C2 is a modification of Algorithm C1 in which the degrees of the nodes are taken into account.  In an attempt to find a larger maximal independent set, the comparison ${deg}(i)$ ${\geq}$ ${deg}(j)$ is used. We will refer to this version of the algorithm as ${{C2_\beta}_0}$. In order to find a smaller maximal independent set, this can be changed to ${deg}(i)$ ${\leq}$ ${deg}(j)$, which will be referred to as ${C2_i}$. The algorithm and proof of correctness are given below.

\begin{algorithm}[htbp]
 \begin{algorithmic}
   \State \textbf{RIn: if} x(i) = 0 ${\wedge}$ (${\nexists}$j ${\in}$ N(i))(x(j) = 1)
   \State ${\;}$ \textbf{then} x(i) = 1

   \State \textbf{ROut: if} x(i) = 1 ${\wedge}$ (${\exists}$j ${\in}$ N(i))(x(j) = 1 ${\wedge}$ ${deg}$(i) ${\geq}$ ${deg}$(j))
   \State ${\;}$ \textbf{then} x(i) = 0
 \end{algorithmic}
 \caption{C2}
 \label{alg:c2}
\end{algorithm}

\begin{lemma} When the system is stable, the set $X$ of nodes $i$ for which ${x(i) = 1}$ forms an independent set.
\label{lem:a2Central}
\end{lemma}
\noindent
\textbf{Proof:} Assume that Algorithm C2 has stabilized, and suppose that the set $X$ is not independent.  Then there are two adjacent nodes ${i}$ and $j$ for which $x(i) = x(j) = 1$.  If the degree of ${i}$ is greater than or equal to the degree of ${j}$, then node $i$ is eligible to execute  \textbf{ROut}. Otherwise, the degree of ${j}$ is greater than the degree of ${i}$, in which case node $j$ is eligible to execute rule \textbf{ROut}. This contradicts the assumption that the system is stable. Therefore, ${X}$ is an independent set.
\qed

\begin{lemma} When the system is stable, the set $X$ forms a maximal independent set.
\label{lem:a2Max}
\end{lemma}
\noindent
\textbf{Proof:} From  Lemma \ref{lem:a2Central}, we know that when the system is stable, $X$ is an independent set. Suppose that $X$ is not a maximal independent set. In that case, there must be a node $i$ not in $X$, none of whose neighbors are in $X$.  In this case, node ${i}$ is eligible to execute rule \textbf{RIn}.  This contradicts the assumption that the system is stable.  Thus,  when the system is stable, the set $X$ is a maximal independent set.
\qed

\begin{lemma} A node ${i}$ that executes \textbf{RIn} will never make another move.
\label{lem:a2RIncomp}
\end{lemma}
\noindent
\textbf{Proof:} Suppose there is a node ${i}$ that executes \textbf{RIn}. By definition, this means that after node $i$ executes \textbf{RIn}, every node $j$ adjacent to $i$ has $x(j) = 0$.  But since $x(i) = 1$, no node adjacent to $i$ can ever execute   \textbf{RIn}.  This, in turn, implies that node $i$ can never become eligible to execute \textbf{ROut}.  Thus, node $i$ will never make another move.
\qed

\begin{lemma} A node ${i}$ that executes \textbf{ROut} can make at most one other move.
\label{lem:a2complexity}
\end{lemma}
\noindent
\textbf{Proof:} Suppose a node ${i}$ executes \textbf{ROut}. Having executed this rule, the  only rule that $i$ can execute is \textbf{RIn}.  But from Lemma 3, it follows that if $i$ executes this rule, it will never make another move..
\qed

\begin{theorem} Algorithm C2 finds a maximal independent set in at most 2n moves.
\label{lem:a2theorem} 
\end{theorem}
\noindent
\textbf{Proof:} See Lemma \ref{lem:a2Max} for correctness and Lemma \ref{lem:a2complexity} for complexity.
\qed

\end{subsection}

\begin{subsection}{Central Scheduler Algorithm 3}\label{CS_3_Subsection}

Algorithm C3 is a modification of Algorithm C2. As with Algorithm C2, ${{C3_\beta}_0}$ and ${C3_i}$ have the comparison signs reversed in order to find a larger and smaller maximal independent set. The idea of Algorithm C3 is that a {\it very weak} vertex, one with a smaller degree than all of its neighbors, can always enter, or be excluded from, the maximal independent set. This is handled by a new rule \textbf{RInVW}. The performance comparison data in \S\ref{Performance_Section} shows that this does produce larger and smaller sets. The algorithm and proof of correctness are given below.

\begin{algorithm}[htbp]
 \begin{algorithmic}

   \State \textbf{RIn: if} x(i) = 0 ${\wedge}$ (${\nexists}$j ${\in}$ N(i))(x(j) = 1)
   \State ${\;}$ \textbf{then} x(i) = 1

   \State \textbf{RInVW: if} x(i) = 0 ${\wedge}$ (${\nexists}$j ${\in}$ N(i))(${deg}$(i) ${\geq}$ ${deg}$(j))
   \State ${\;}$ \textbf{then} x(i) = 1

   \State \textbf{ROut: if} x(i) = 1 ${\wedge}$ (${\exists}$j ${\in}$ N(i))(x(j) = 1 ${\wedge}$ ${deg}$(i) ${\geq}$ ${deg}$(j))
   \State ${\;}$ \textbf{then} x(i) = 0
 \end{algorithmic}
 \caption{C3}
 \label{alg:c3}
\end{algorithm}

\begin{lemma} When the system is stable, the set $X$ forms a maximal independent set.
\label{lem:a3Max}
\end{lemma}
\noindent
\textbf{Proof:} With a slight modification, this can be seen from Lemma \ref{lem:a2Max}.
\qed

\begin{lemma} A node ${i}$ that executes \textbf{RInVW} will never make another move.
\label{lem:a3RInVW}
\end{lemma}
\noindent
\textbf{Proof:} Suppose a node ${i}$ does make another move after executing \textbf{RInVW}. In this case, the next move must be \textbf{ROut} and ${i}$ must have a neighbor ${j \in X}$. However, for ${i}$ to have executed \textbf{RInVW}, it must have a smaller degree than all of its neighbors. This violates the condition of \textbf{ROut} that a neighbor in ${X}$ has a degree equal to, or smaller, than ${i}$. Therefore, a node ${i}$ will not make another move after executing \textbf{RInVW}.
\qed

\begin{lemma} A node ${i}$ that executes \textbf{RIn} will make at most one more move.
\label{lem:a3RIn}
\end{lemma}
\noindent
\textbf{Proof:} For ${i}$ to make another move, it must next execute \textbf{ROut}. Additionally, for ${i}$ to have executed \textbf{RIn}, none of its neighbors could have been in $X$ at the time. Therefore, a neighbor of ${i}$ must be a {\it very weak} node and have executed \textbf{RInVW}. As seen from Lemma \ref{lem:a3RInVW}, this node will never make another move, so after ${i}$ executes \textbf{ROut}, it will be next to a {\it very weak} node in $X$ and cannot execute any further rules.
\qed

\begin{lemma} A node ${i}$ that executes \textbf{ROut} will make at most two more moves.
\label{lem:a3ROut}
\end{lemma}
\noindent
\textbf{Proof:} Assume a node ${i \in X}$ is next to a neighbor ${j \in X}$ from the initial configuration. Node ${i}$ can then execute \textbf{ROut}. From Lemma \ref{lem:a3RIn}, node ${i}$ can execute at most two more moves.
\qed

\begin{theorem} Algorithm C3 finds a maximal independent set in at most 3n moves.
\label{lem:a3theorem} 
\end{theorem}
\noindent
\textbf{Proof:} See Lemma \ref{lem:a3Max} for correctness and Lemmas \ref{lem:a3RInVW} and \ref{lem:a3RIn}, and \ref{lem:a3ROut} for complexity.
\qed

\end{subsection}

\begin{subsection}{Central Scheduler Algorithm 4}\label{CS_4_Subsection}

Algorithm C4 is a further modification of Algorithm C3. As with Algorithm C2 and C3, ${{C4_\beta}_0}$ and ${C4_i}$ have the comparison signs reversed in order to find a larger and smaller maximal independent set. Algorithm C3 was concerned with {\it very weak} vertices. Algorithm C4 relaxes this to examine {\it relatively very weak} vertices. A vertex is {\it relatively very weak} if it has a smaller degree than all of its neighbors that are in the set ${X}$.

\begin{algorithm}[htbp]
 \begin{algorithmic}
   \State \textbf{RIn: if} x(i) = 0 ${\wedge}$ (${\nexists}$j ${\in}$ N(i))(x(j) = 1 ${\wedge}$ ${deg}$(i) ${\geq}$ ${deg}$(j))
   \State ${\;}$ \textbf{then} x(i) = 1

   \State \textbf{ROut: if} x(i) = 1 ${\wedge}$ (${\exists}$j ${\in}$ N(i))(x(j) = 1 ${\wedge}$ ${deg}$(i) ${\geq}$ ${deg}$(j))
   \State ${\;}$ \textbf{then} x(i) = 0
 \end{algorithmic}
 \caption{C4}
 \label{alg:c4}
\end{algorithm}

A solution was developed independently by Yen, Huang, and Turau in \cite{vt16} for a similar algorithm named $\mathcal{A}_{deg}$, which provides an approximation ratio that was bounded by $(\Delta + 2) / 3$. Their solution utilizes game theory and provides several measurements for its performance compared to other algorithms. Their findings also correspond with ours in Section \ref{Performance_Section} that this approach, considering the degree of the node, provides the best solution for maximal independent sets.

\end{subsection}

\begin{subsection}{Central Scheduler Algorithm 5}\label{CS_5_Subsection}

Algorithm C5 is modeled after the Grundy coloring algorithm of Hedetniemi, Jacobs and Srimani \cite{Hedet}, and is found in \cite{KKen} by Hedetniemi, Jacobs and Kennedy as an example of a self-stabilizing algorithm that can find two disjoint independent sets. In Algorithm C5, every node is assigned a color ${x(i) \in \{1, 2, 3\}}$ where:

\begin{equation}
  {R = \{i{\mid}x(i) = 1\}}
\end{equation}
\begin{equation}
  {B = \{i{\mid}x(i) = 2\}}
\end{equation}

and the following function is defined: 

\begin{center}
Color(i) = min(\{1, 2, 3\} - (\{x(j) $\mid$ j $\in$ N(i)\} $\cap$ \{1, 2\}))
\end{center}

When the algorithm stabilizes, as shown in \cite{KKen}, $R$ is a maximal independent set in $G = (V,E)$, and $B$ is a  maximal independent set in the induced subgraph $G[V - R] = (V - R, (V-R \times V-R) \cap E)$.

\begin{algorithm}[htbp]
 \begin{algorithmic}
   \State \textbf{Re-color: if} x(i) $\neq$ Color(i)
   \State ${\;}$ \textbf{then} x(i) = Color(i)

 \end{algorithmic}
 \caption{C5}
 \label{alg:c5}
\end{algorithm}

\begin{theorem} Algorithm C5 finds a maximal independent set $R$ in $G$ and $B$ in $G[V - R]$ in at most 3n moves.
\label{lem:c5theorem} 
\end{theorem}
\noindent
\textbf{Proof:} See proof of Algorithm 3 in \cite{KKen}.
\qed

\end{subsection}

\end{section}



\begin{section}{Distributed Algorithms}\label{DS_Section}

In this section, we present five self-stabilizing algorithms for finding maximal independent sets running under the distance-1 model with an unfair distributed scheduler, whereby any number of eligible nodes may move at the same time.  The algorithms are all silent. Four algorithms are new and proofs are given for their correctness and complexity. As before, results are given for the other algorithm with the reader being referred to the original paper for its proof. For each of the algorithms, we assume that the nodes $i$ have unique identifiers $id(i)$. 

\begin{subsection}{Distributed Scheduler Algorithm 1}\label{DS_1_Subsection}

Algorithm D1 was designed by Turau (see \cite{vt07}) as a generalization of Algorithm C1. The algorithm includes ${wait}$ and ${back}$ states to take into account cases in which two adjacent nodes are eligible to move at the same time. In case two adjacent nodes want to enter the independent set $X$ at once by executing rule \textbf{RIn}, ids are used to determine which node will be allowed to enter the set. 

\begin{algorithm}[htbp]
 \begin{algorithmic}
   \State \textbf{RWait: if} x(i) = 0 ${\wedge}$ (${\nexists}$j ${\in}$ N(i))(x(j) = 1)
   \State ${\;}$ \textbf{then} x(i) = ${w}$
   \State \textbf{RBack: if} x(i) = ${w}$ ${\wedge}$ (${\exists}$j ${\in}$ N(i))(x(j) = 1)
   \State ${\;}$ \textbf{then} x(i) = 0
   \State \textbf{RIn: if} x(i) = ${w}$ ${\wedge}$ (${\nexists}$j ${\in}$ N(i))(x(j) = 1) ${\wedge}$ (${\forall}$k ${\in}$ N(i))(x(k) ${\neq}$ ${w}$ ${\vee}$ id(k) ${>}$ id(i))
   \State ${\;}$ \textbf{then} x(i) = 1
   \State \textbf{ROut: if} x(i) = 1 ${\wedge}$ (${\exists}$j ${\in}$ N(i))(x(j) = 1)
   \State ${\;}$ \textbf{then} x(i) = 0
 \end{algorithmic}
 \caption{D1}
 \label{alg:d1}
\end{algorithm}

\begin{theorem} Algorithm D1 finds a maximal independent set in at most ${max(3n - 5, 2n)}$ moves.
\label{lem:d1theorem} 
\end{theorem}
\noindent
\textbf{Proof:} See proof of Algorithm ${\mathcal{A}_{\text{MIS}}}$ in \cite{vt07}.
\qed

\end{subsection}

\begin{subsection}{Distributed Scheduler Algorithm 2}\label{DS_2_Subsection}

Algorithm D2 is the distributed version of Algorithm C2, the new algorithm presented earlier for finding a maximal independent set. As with Algorithm C2, the comparison ${deg}(i)$ ${\geq}$ ${deg}(j)$ can be changed to ${deg}(i)$ ${\leq}$ ${deg}(j)$ to find a smaller maximal independent set. This change will be referred to as Algorithm D2-1. 

\begin{algorithm}[htbp]
 \begin{algorithmic}
    \State \textbf{RWait: if} x(i) = 0 ${\wedge}$ (${\nexists}$j ${\in}$ N(i))(x(j) = 1)
   \State ${\;}$ \textbf{then} x(i) = ${w}$
   \State \textbf{RBack: if} x(i) = ${w}$ ${\wedge}$ (${\exists}$j ${\in}$ N(i))(x(j) = 1)
   \State ${\;}$ \textbf{then} x(i) = 0
   \State \textbf{RIn: if} x(i) = ${w}$ ${\wedge}$ (${\nexists}$j ${\in}$ N(i))(x(j) = 1) ${\wedge}$ (${\forall}$k ${\in}$ N(i))(x(k) ${\neq}$ ${w}$ ${\vee}$ id(k) ${\geq}$ id(i))
   \State ${\;}$ \textbf{then} x(i) = 1
   \State \textbf{ROut: if} x(i) = 1 ${\wedge}$ (${\exists}$j ${\in}$ N(i))(x(j) = 1 ${\wedge}$ ${deg}$(i) ${\geq}$ ${deg}$(j))
   \State ${\;}$ \textbf{then} x(i) = 0
 \end{algorithmic}
 \caption{D2}
 \label{alg:d2}
\end{algorithm}

\begin{lemma} When the system is stable, the set $X$ of nodes with $x(i) = 1$ forms an independent set.
\label{lem:d2IndeCorrect}
\end{lemma}
\noindent
\textbf{Proof:} See proof of Lemma \ref{lem:a2Central} with slight modification for wait state and ids.
\qed

\begin{lemma}
When the system is stable, no node $i$ has $x(i) = w$.
\end{lemma}
\noindent
\textbf{Proof:}  Assume that the system is stable and there exists a node $i$ with $x(i) = w$.  Assume furthermore, that node $i$ has the smallest $id$ of all nodes $j$ with $x(j) = w$.  Since the system is stable, node $i$ is not eligible to execute rule \textbf{RBack}.  This implies that no neighbor $j \in N(i)$ has $x(j) = 1$.  Thus, every neighbor $j \in N(i)$ has $x(j) \in \{0, w\}$.  But since the system is stable, and node $i$ is not eligible to execute rule \textbf{RIn}, node $i$ must have at least one neighbor, say $k \in N(i)$, with $x(k) = w$ and $id(k) < id(i)$.  But this contradicts our assumption that node (i) has the smallest $id$ among all nodes with $x(j) = w$.
Thus, when the system is stable, no node $i$ has $x(i) = w$.
\qed

\begin{lemma} When the system is stable, the set $X$ forms a maximal independent set.
\label{lem:d2MaxIndCorrect}
\end{lemma}
\noindent
\textbf{Proof:} Assume that the system is stable and the set $X$ does not form a maximal independent set.  From Lemma \ref{lem:d2IndeCorrect}, we know that when the system is stable $X$ is an independent set. Therefore, $X$ is not a maximal independent set.  This implies that there exists a node $i$ with $x(i) \in \{ 0, w\}$ and no neighbor $j \in N(i)$ has $x(j) = 1$.  But from Lemma 6, we know that if the system is stable, then no node $i$ has $x(i) = w$.  Therefore, there must exist a node $i$ with $x(i) = 0$ and all neighbors $j \in N(i)$ have $x(j) = 0$.  Therefore, node (i) is eligible to execute rule \textbf{RWait}, and the system is not stable.
\qed

\begin{lemma} If a node i executes rule \textbf{RIn}, it will never make another move.
\label{lem:d2RInComplex}
\end{lemma}
\noindent
\textbf{Proof:} Assume that a node $i$ executes rule \textbf{RIn}.  The only rule it could be eligible to execute after this is rule \textbf{ROut}. By contradiction, suppose that node ${i}$ executes \textbf{ROut} sometime after executing \textbf{RIn}. Therefore, a neighbor $j$ of ${i}$ must have entered the independent set after, or at the same time, as ${i}$. From Lemma~\ref{lem:a2RIncomp}, we know that a neighbor ${j}$ of ${i}$ could not have executed \textbf{RIn} while ${i}$ is in the independent set. Therefore, ${j}$ must have executed \textbf{RIn} at the same time as ${i}$. However, the guard of \textbf{RIn} requires that a node either have no neighbors in the wait state or any neighbor in the wait state must have a larger id. As the ids are unique and ${i}$ executed \textbf{RIn}, node ${j}$ must have a larger id than ${i}$ and could not enter the independent set. Thus a node ${i}$ will never make another move after executing \textbf{RIn}.
\qed

\begin{lemma} A node i that executes rule \textbf{RWait} can make at most one more move.
\label{lem:d2RWaitComplex}
\end{lemma}
\noindent
\textbf{Proof:} Suppose there is a node ${i}$ that executes \textbf{RWait}. Node ${i}$ is now in the wait state and can only execute \textbf{RBack} or \textbf{RIn}. 

\textbf{Case 1:} node ${i}$ executes \textbf{RIn}. From lemma \ref{lem:d2RInComplex}, we see that node ${i}$ will never make another move. 

\textbf{Case 2:} node ${i}$ executes \textbf{RBack}. For node ${i}$ to have executed \textbf{RBack}, a node $j \in N(i)$ must have executed \textbf{RIn} in which case node ${i}$ will not be able to execute another move.
\qed

\begin{lemma} A node ${i}$ that executes \textbf{ROut} will make at most two more moves.
\label{lem:d2ROutComplex}
\end{lemma}

\noindent
\textbf{Proof:} Suppose there is a node ${i}$ that executes \textbf{ROut} from its initial state. For node ${i}$ to have executed \textbf{ROut}, there had to have been a neighbor ${j}$ such that ${x(j) = 1}$ and ${deg(i) \geq deg(j)}$. It is possible that node ${j}$ executes \textbf{ROut} at the same time as ${i}$. Thus, node ${i}$ could execute \textbf{RWait}, and from Lemma \ref{lem:d2RWaitComplex}, we see that ${i}$ will make at most one more move.  
\qed

\begin{lemma} A node ${i}$ that executes \textbf{RBack} will make at most two more moves.
\label{lem:d2RBackComplex}
\end{lemma}

\noindent
\textbf{Proof:} Suppose there is a node ${i}$ that executes \textbf{RBack} from its initial state. For node ${i}$ to have executed \textbf{RBack}, there had to have been a neighbor ${j}$ such that ${x(j) = 1}$. If node ${j}$ then executes \textbf{ROut} (only from the initial state), the next move that ${i}$ can make is \textbf{RWait}, and from Lemma~\ref{lem:d2RWaitComplex}, we see that ${i}$ could make at most one more move.
\qed

\begin{theorem} Algorithm D2 finds a maximal independent set in at most 3n moves.
\label{lem:d2theorem} 
\end{theorem}
\noindent
\textbf{Proof:} This follows immediately from Lemmas \ref{lem:d2MaxIndCorrect}, \ref{lem:d2ROutComplex} and \ref{lem:d2RBackComplex}.
\qed

\end{subsection}

\begin{subsection}{Distributed Scheduler Algorithm 3}\label{DS_3_Subsection}

\begin{algorithm}[htbp]
 \begin{algorithmic}
    \State \textbf{RWait: if} x(i) = 0 ${\wedge}$ ((${\nexists}$j ${\in}$ N(i))(x(j) = 1) ${\vee}$ (${\nexists}$k ${\in}$ N(i))(${deg}$(i) ${\geq}$ ${deg}$(k)))

   \State ${\;}$ \textbf{then} x(i) = ${w}$
   \State \textbf{RBack: if} x(i) = ${w}$ ${\wedge}$ (${\exists}$j ${\in}$ N(i))(x(j) = 1) ${\wedge}$ (${\exists}$k ${\in}$ N(i))(${deg}$(i) ${\geq}$ ${deg}$(k))
   \State ${\;}$ \textbf{then} x(i) = 0

   \State \textbf{RIn: if} x(i) = ${w}$ ${\wedge}$ (${\nexists}$j ${\in}$ N(i))(x(j) = 1) ${\wedge}$ (${\forall}$k ${\in}$ N(i))(x(k) ${\neq}$ ${w}$ ${\vee}$ id(k) ${\geq}$ id(i))
   \State ${\;}$ \textbf{then} x(i) = 1

   \State \textbf{RInVW: if} x(i) = ${w}$ ${\wedge}$ (${\nexists}$j ${\in}$ N(i))(${deg}$(i) ${\geq}$ ${deg}$(j))
   \State ${\;}$ \textbf{then} x(i) = 1

   \State \textbf{ROut: if} x(i) = 1 ${\wedge}$ (${\exists}$j ${\in}$ N(i))(x(j) = 1 ${\wedge}$ ${deg}$(i) ${\geq}$ ${deg}$(j))
   \State ${\;}$ \textbf{then} x(i) = 0
 \end{algorithmic}
 \caption{D3}
 \label{alg:d3}
\end{algorithm}

\begin{lemma} When the system is stable, the set $X$ forms a maximal independent set.
\label{lem:d3MaxIndCorrect}
\end{lemma}
\noindent
\textbf{Proof:} This can be seen with a slight modification of Lemma \ref{lem:d2MaxIndCorrect}.
\qed

\begin{lemma} A node ${i}$ that executes \textbf{RInVW} will never make another move.
\label{lem:d3RInVW}
\end{lemma}
\noindent
\textbf{Proof:} See proof of Lemma \ref{lem:a3RInVW}.
\qed

\begin{lemma} A node ${i}$ that executes \textbf{RIn} will make at most one other move.
\label{lem:d3RIn}
\end{lemma}
\noindent
\textbf{Proof:} From Lemmas \ref{lem:d2RInComplex} and \ref{lem:d3RInVW}, if node ${i}$ executes \textbf{RIn}, the only possible move for node ${i}$ to make will be to execute \textbf{ROut}. This can only occur if a neighbor ${j \in N(i)}$ executes \textbf{RInVW}. At that point, node ${i}$ will not make another move as \textbf{RWait} will not be enabled.
\qed

\begin{lemma} A node i that executes rule \textbf{RWait} can make at most 2 * deg(i) - 1 more moves.
\label{lem:d3RWaitComplex}
\end{lemma}
\noindent
\textbf{Proof:} Suppose there is a node ${i}$ that executes \textbf{RWait}. Node ${i}$ is now in the wait state and can execute \textbf{RBack}, \textbf{RIn}, or \textbf{RInVW}.

\textbf{Case 1:} node ${i}$ executes \textbf{RInVW}. From lemma \ref{lem:d3RInVW}, we see that node ${i}$ will never make another move.

\textbf{Case 2:} node ${i}$ executes \textbf{RIn}. From lemma \ref{lem:d3RIn}, we see that node ${i}$ will make at most one more move.

\textbf{Case 3:} node ${i}$ executes \textbf{RBack}. For node ${i}$ to have executed \textbf{RBack}, a neighbor $j \in N(i)$ must have executed \textbf{RIn} or \textbf{RInVW}. From Lemma \ref{lem:d3RInVW}, if node ${j}$ executes \textbf{RInVW}, ${j}$ will not make another move nor will node ${i}$. However from Lemma \ref{lem:d3RIn}, if node ${j}$ executed \textbf{RIn}, a neighbor ${k \in N(j)}$ where ${k \neq i}$ may execute \textbf{RInVW}. After such a move, node ${j}$ must execute \textbf{ROut} allowing ${i}$ to once again execute \textbf{RWait}. This set of moves can occur up to $deg(i)$ times allowing for a maximum of $2 * deg(i) - 1$ moves at node ${i}$ after $i$ has executed the initial \textbf{RWait}.
\qed

\begin{lemma} A node i that executes rule \textbf{RBack} can make at most 2 * deg(i) more moves.
\label{lem:d3RBackComplex}
\end{lemma}
\noindent
\textbf{Proof:} A node ${i}$ may initially be in the wait state with a neighbor ${j \in N(i)}$ initially in the set X. Node $i$ can then execute \textbf{RBack}. The next move node $i$ could make is to execute \textbf{RWait}, at which point from Lemma \ref{lem:d3RWaitComplex}, node $i$ can make at most $2 * deg(i) - 1$ more moves.
\qed

\begin{lemma} A node i may execute at most 2 \textbf{ROut} moves.
\label{lem:d3ROut}
\end{lemma}

\noindent
\textbf{Proof:} Nodes $i$ and $j$ where${j \in N(i)}$ may be initially in the set X. At this point, $i$ can execute \textbf{ROut}. From Lemma~\ref{lem:d3RInVW}, if a node executes \textbf{RInVW}, it will never make another move, and from Lemma~\ref{lem:d3RIn}, if a node executes \textbf{RIn}, it will execute \textbf{ROut} at most one more time.
\qed

\begin{theorem} Algorithm D3 finds a maximal independent set in at most O(${\Delta}$n) moves.
\label{lem:d3theorem}
\end{theorem}
\noindent
\textbf{Proof:} See Lemma \ref{lem:d3MaxIndCorrect} for correctness and Lemmas \ref{lem:d3RInVW}, \ref{lem:d3RIn}, \ref{lem:d3RWaitComplex}, \ref{lem:d3RBackComplex}, and \ref{lem:d3ROut} for complexity.
\qed

\end{subsection}

\begin{subsection}{Distributed Scheduler Algorithm 4}\label{DS_4_Subsection}

\begin{algorithm}[htbp]
 \begin{algorithmic}
    \State \textbf{RWait: if} x(i) = 0 ${\wedge}$ ((${\nexists}$j ${\in}$ N(i))(x(j) = 1) ${\vee}$ (${\nexists}$k ${\in}$ N(i))(x(k) = 1 ${\wedge}$ ${deg}$(i) ${\geq}$ ${deg}$(k)))

   \State ${\;}$ \textbf{then} x(i) = ${w}$
   \State \textbf{RBack: if} x(i) = ${w}$ ${\wedge}$ (${\exists}$j ${\in}$ N(i))(x(j) = 1) ${\wedge}$ (${\exists}$k ${\in}$ N(i))(x(k) = 1 ${\wedge}$ ${deg}$(i) ${\geq}$ ${deg}$(k))
   \State ${\;}$ \textbf{then} x(i) = 0

   \State \textbf{RIn: if} x(i) = ${w}$ ${\wedge}$ (${\nexists}$j ${\in}$ N(i))(x(j) = 1 ${\wedge}$ ${deg}$(i) ${\geq}$ ${deg}$(j))
   \State ${\;}$ \textbf{then} x(i) = 1

   \State \textbf{ROut: if} x(i) = 1 ${\wedge}$ (${\exists}$j ${\in}$ N(i))(x(j) = 1 ${\wedge}$ ${deg}$(i) ${\geq}$ ${deg}$(j))
   \State ${\;}$ \textbf{then} x(i) = 0
 \end{algorithmic}
 \caption{D4}
 \label{alg:d4}
\end{algorithm}

The maximum number of moves for this algorithm remains an open problem. As stated for Algorithm C4, an approximation ratio was developed by Yen, Huang, and Turau in \cite{vt16} showing that the central solution was bounded by $(\Delta + 2) / 3$.

\end{subsection}

\begin{subsection}{Distributed Scheduler Algorithm 5}\label{DS_5_Subsection}

Algorithm D5, a distributed version of Algorithm C5, is given below. In \cite{KKen}, no distributed version of Algorithm C5 was given. We have included one below along with a proof of its correctness and complexity. $R$, $B$, and $Color$ are defined in \S\ref{CS_5_Subsection}. Additionally, we define the following function:

\begin{center}
swn($i$) $\equiv$ $\exists$j $\in$ N(i)(id(j) $<$ id(i) $\wedge$ x(j) = w)
\end{center}

\begin{algorithm}[htbp]
 \begin{algorithmic}
   \State \textbf{RWait: if} x(i) ${=}$ 3 ${\wedge}$ x(i) ${\neq}$ Color(i)
   \State ${\;}$ \textbf{then} x(i) = ${w}$

   \State \textbf{RBack: if} x(i) = ${w}$ ${\wedge}$ Color(i) = 3
   \State ${\;}$ \textbf{then} x(i) = 3

   \State \textbf{RIn: if} x(i) = ${w}$ ${\wedge}$ Color(i) ${\neq}$ 3 ${\wedge}$ ${\neg}$swn(i)
   \State ${\;}$ \textbf{then} x(i) = Color(i)

   \State \textbf{ROut: if} x(i) ${\in}$ \{1, 2\} ${\wedge}$ x(i) ${\neq}$ Color(i)
   \State ${\;}$ \textbf{then} x(i) = 3
 \end{algorithmic}
 \caption{D5}
 \label{alg:d5}
\end{algorithm}

\begin{lemma} When the system is stable, no node is in the wait state.
\label{lem:d5WaitSet}
\end{lemma}

\noindent
\textbf{Proof:} By contradiction, let us assume that the system is stable and there is a node ${i}$ in the wait state that has the smallest ID of all nodes in the wait state. In this case, node $i$ is eligible to execute \textbf{RBack} or \textbf{RIn}, and therefore, the system is not stable.\qed

\begin{lemma} When the system is stable, ${R}$ is maximal independent in ${G}$, and ${B}$ is maximal independent in $G[V-R]$.
\label{lem:d5MaxSet}
\end{lemma}

\noindent
\textbf{Proof:} By Lemma \ref{lem:d5WaitSet}, when the system is stable, no node is in the wait state. The set ${R}$ must be an independent set; otherwise, two nodes $i, j \in R$ would be adjacent having $x(i) = x(j) = 1$ and both would be eligible to execute \textbf{ROut}. A similar argument shows that ${B}$ must be an independent set. Additionally, ${R}$ must be maximal independent in ${G}$. If a node ${j}$ is not adjacent to a node in ${R}$, it can execute either \textbf{RWait} or \textbf{RIn}. Likewise, ${B}$ must be maximal independent in $G[V-R]$.
\qed

\begin{lemma} If a node i executes \textbf{RIn} and x(i) = 1, then it will never make another move.
\label{lem:d5RIn}
\end{lemma}

\noindent
\textbf{Proof:} Due to the $swn$ function, we know that all neighbors of ${i}$ in the wait state must have larger $id$s than ${i}$. Additionally, from the $Color$ function, no neighbor of ${i}$ can have the color 1. Therefore, after ${i}$ executes \textbf{RIn}, no neighbor of ${i}$ will be able to set its color to 1, and ${i}$ will never move again.
\qed

\begin{lemma} If a node i executes \textbf{RIn} and x(i) = 2, then node i will make at most 2 * deg(i) - 1 more moves. 
\label{lem:d5RInTwo}
\end{lemma}

\noindent
\textbf{Proof:} Suppose node ${i}$ has the largest id of all of its neighbors and executes rule \textbf{RIn} and x(i) = 2. As in Lemma~\ref{lem:d5RIn}, no neighbor of ${i}$ will set its color to 2 while ${i}$ has color 2. Therefore, for ${i}$ to move again, a neighbor ${j}$ with $x(j) = 1$ must execute \textbf{ROut}. This is possible in the initial state if ${j}$ is adjacent to a neighbor also colored 1. At this point, ${i}$ can execute \textbf{ROut} and then \textbf{RWait} as it has no neighbors colored 1.

\textbf{Case 1:} If ${i}$ next executes \textbf{RIn} and sets $x(i) = 1$, from Lemma~\ref{lem:d5RIn}, we know that ${i}$ will never make another move. 

\textbf{Case 2:} A neighbor ${k}$ of ${i}$ sets $x(k) = 2$. Node ${i}$ can now execute \textbf{RBack}. Assume ${k}$ is also adjacent to a neighbor ${l}$ with $x(l) = 1$ from the initial state. As with ${j}$, ${l}$ may execute \textbf{ROut} allowing ${k}$ to set its color to 1. At this point, node ${i}$ can again execute \textbf{RWait} to attempt to set $x(i) = 2$. This process can repeat for $deg(i)$ - 1 times. At the end, ${i}$ will either execute \textbf{RBack} or \textbf{RIn}.
\qed

\begin{lemma} A node i may execute at most deg(i) \textbf{RWait} moves.
\label{lem:d5RWait}
\end{lemma}

\noindent
\textbf{Proof:} This follows from Lemma~\ref{lem:d5RInTwo}.
\qed

\begin{lemma} A node i may execute at most deg(i) + 1 \textbf{RBack} moves.
\label{lem:d5RBack}
\end{lemma}

\noindent
\textbf{Proof:} A node ${i}$ may initially be in the wait state when it has neighbors colored 1 and 2. At this point, ${i}$ may execute \textbf{RBack}. From Lemma~\ref{lem:d5RWait}, we see that there may be $deg(i)$ \textbf{RWait} moves, each of which may be followed by an \textbf{RBack} move.
\qed

\begin{lemma} A node i may execute at most 2 \textbf{ROut} moves.
\label{lem:d5ROut}
\end{lemma}

\noindent
\textbf{Proof:} A node in the initial state with color 1 or 2 may execute \textbf{ROut} if it is adjacent to a neighbor with the same color. From Lemma~\ref{lem:d5RIn}, we know that after a node executes \textbf{RIn} and sets $x(i) = 1$, it will never move again. However, if it sets ${x(i) = 2}$, it may execute one more \textbf{ROut} rule if a neighbor ${j}$ where $x(j) = 1$ from the initial configuration executes \textbf{ROut}.
\qed

\begin{theorem} Algorithm D5 finds a maximal independent set ${R}$ in ${G}$ and a maximal independent set ${B}$ in $G[V-R]$ in $O({\Delta}n)$ moves.
\label{lem:d5theorem}
\end{theorem}
\noindent
\textbf{Proof:} See Lemma \ref{lem:d5MaxSet} for correctness and Lemmas \ref{lem:d5RIn}, \ref{lem:d5RInTwo}, \ref{lem:d5RWait}, \ref{lem:d5RBack}, and \ref{lem:d5ROut} for complexity.
\qed

\end{subsection}

\end{section}


\begin{section}{Performance Comparisons}\label{Performance_Section}
In the previous two sections, we have presented several self-stabilizing algorithms for finding maximal independent sets, using both central and distributed schedulers. We implemented each algorithm and run simulations on four classes of graphs: trees (connected and acyclic graphs), bipartite graphs (connected graphs having no cycles of odd length), unit disk graphs (the intersection graphs of unit disks randomly placed in the plane) and arbitrary connected graphs.  In each case we randomly generated 5,000 graphs of order $n = 500$, for each of these four types.  For algorithms using the central scheduler, we selected at random any of the currently eligible nodes to move, and for algorithms using the distributed scheduler, we randomly selected a subset of the set of currently eligible nodes to move.  In each case we recorded the number of moves made before the algorithm stabilized, and the size of the independent set found by the algorithm.


\begin{table*}[ht]
\centering
\begin{tabular}{c | c c | c c | c c | c c}
\hline
Algorithm & Tree & \% Diff. & Bipartite & \% Diff. & Unit Disk & \% Diff. & Connected & \% Diff. \\
\hline
${C1}$ & 267.5 & - & 167.6 & - & 61.1 & - & 95.1 & - \\
${{C2_\beta}_0}$ & 282.5 & 5.5\% & 178.5 & 6.3\% & 65.9 & 7.6\% & 102.8 & 7.8\% \\
${C2_i}$ & 247.3 & -7.8\% & 154.4 & -8.2\% & 58.1 & -5.0\% & 86.5 & -9.5\% \\
${{C3_\beta}_0}$ & 296.9 & 10.4\% & 185.7 & 10.2\% & 68.0 & 10.7\% & 106.4 & 11.2\% \\
${C3_i}$ & 232.5 & -14.0\% & 149.9 & -11.1\% & 57.6 & -5.9\% & 85.0 & -11.2\% \\
\rowcolor{table-gray}
${{C4_\beta}_0}$ & 297.2 & 10.5\% & 186.5 & 10.7\% & 71.3 & 15.4\% & 107.6 & 12.3\% \\
\rowcolor{table-gray}
${C4_i}$ & 221.1 & -19.0\% & 148.1 & -12.4\% & 57.2 & -6.6\% & 80.4 & -16.8\% \\
${C5}$ & 267.2 & -0.1\% & 167.3 & -0.2\% & 61.1 & 0.0\% & 95.0 & -0.1\% \\
\hline
${D1}$ & 264.2 & - & 166.7 & - & 60.9 & - & 94.4 & - \\
${{D2_\beta}_0}$ & 272.8 & 3.2\% & 174.3 & 4.5\% & 64.5 & 5.7\% & 100.6 & 6.4\% \\
${D2_i}$ & 253.2 & -4.3\% & 157.7 & -5.5\% & 58.4 & -4.2\% & 87.9 & -7.1\% \\
${{D3_\beta}_0}$ & 296.7 & 11.6\% & 185.8 & 10.8\% & 67.5 & 10.3\% & 106.3 & 11.9\% \\
${D3_i}$ & 233.3 & -12.4\% & 150.8 & -10.0\% & 57.8 & -5.2\% & 85.6 & -9.8\% \\
\rowcolor{table-gray}
${{D4_\beta}_0}$ & 297.1 & 11.7\% & 186.4 & 11.2\% & 71.3 & 15.7\% & 107.6 & 13.1\% \\
\rowcolor{table-gray}
${D4_i}$ & 221.1 & -17.8\% & 147.5 & -12.2\% & 57.2 & -6.3\% & 80.4 & -16.0\% \\
${D5}$ & 263.3 & -0.3\% & 166.3 & -0.2\% & 60.9 & 0.0\% & 94.2 & -0.2\% \\
\hline
\end{tabular}
\caption{Average size of maximal independent sets found}
\label{tab:tSetSize}
\end{table*}



\begin{table*}[ht]
\centering
\begin{tabular}{c | c | c c | c c | c c | c c }
\hline
Algorithm & Complexity & Tree & \% Diff. & Bipartite & \% Diff. & Unit Disk & \% Diff. & Connected & \% Diff. \\
\hline
${C1}$ & $2n$ & 205.4 & - & 253.5 & - & 239.1 & - & 275.9 & - \\
\rowcolor{table-gray}
${{C2_\beta}_0}$ & $2n$ & 182.0 & -12.1\% & 200.8 & -23.2\% & 215.4 & -10.4\% & 225.7 & -20.0\% \\
\rowcolor{table-gray}
${C2_i}$ & $2n$ & 199.7 & -2.8\% & 240.0 & -5.5\% & 237.5 & -0.7\% & 259.3 & -6.2\% \\
${{C3_\beta}_0}$ & $3n$ & 274.0 & 28.6\% & 259.6 & 2.4\% & 237.9 & -0.5\% & 261.1 & -5.5\% \\
${C3_i}$ & $3n$ & 253.3 & 20.9\% & 274.3 & 7.9\% & 246.9 & 3.2\% & 276.2 & 0.1\% \\
${{C4_\beta}_0}$ & ? & 285.8 & 32.7\% & 288.2 & 12.8\% & 300.0 & 22.6\% & 302.8 & 9.3\% \\
${C4_i}$ & ? & 389.7 & 61.9\% & 396.6 & 44.0\% & 344.9 & 36.2\% & 382.5 & 32.4\% \\
${C5}$ & $3n$ & 302.4 & 38.2\% & 331.8 & 26.8\% & 307.5 & 25.0\% & 372.0 & 29.7\% \\
\hline
${D1}$ & ${max}$(3${n}$-5,2${n}$) & 388.1 & - & 422.2 & - & 393.8 & - & 474.6 & - \\
\rowcolor{table-gray}
${{D2_\beta}_0}$ & $3n$ & 383.4 & -1.2\% & 381.8 & -10.0\% & 351.8 & -11.3\% & 389.5 & -19.7\% \\
\rowcolor{table-gray}
${D2_i}$ & $3n$ & 376.9 & -2.9\% & 394.5 & -6.8\% & 369.3 & -6.4\% & 416.7 & -13.0\% \\
${{D3_\beta}_0}$ & $O({\Delta}n)$ & 490.1 & 23.2\% & 449.8 & 6.3\% & 387.1 & -1.7\% & 429.2 & -10.0\% \\
${D3_i}$ & $O({\Delta}n)$ & 444.7 & 13.6\% & 452.8 & 7.0\% & 388.9 & -1.3\% & 450.4 & -5.2\% \\
${{D4_\beta}_0}$ & ? & 511.8 & 27.5\% & 487.7 & 14.4\% & 481.1 & 20.0\% & 484.6 & 2.1\% \\
${D4_i}$ & ? & 659.6 & 51.8\% & 655.8 & 43.3\% & 550.9 & 33.3\% & 624.8 & 27.3\% \\
${D5}$ & $O({\Delta}n)$ & 646.1 & 49.9\% & 628.6 & 39.3\% & 513.4 & 26.4\% & 688.6 & 36.8\% \\
\hline
\end{tabular}
\caption{Worst-case complexity and average number of moves}
\label{tab:tCompSteps}
\end{table*}

\begin{figure*}
  \begin{tikzpicture}
    \begin{axis}[
      xlabel=Order $n$,
      ylabel=Number of Moves,
      legend pos=outer north east]
      \addplot[smooth,mark=*,color=dark-gray]
        plot coordinates {
        (250,143.2)
        (500,285.8)
        (1000,570.7)
        (2000,1143.9)
      };
      \addlegendentry{${{C4_\beta}_0}$}

      \addplot[smooth,color=dark-gray,mark=x]
        plot coordinates {
        (250,193.9)
        (500,389.7)
        (1000,785.6)
        (2000,1561.1)
      };
      \addlegendentry{${C4_i}$}

      \addplot[smooth,color=dark-gray,mark=square]
        plot coordinates {
        (250,255.9)
        (500,511.8)
        (1000,1024.6)
        (2000,2046.0)
      };
      \addlegendentry{${{D4_\beta}_0}$}

      \addplot[smooth,color=dark-gray,mark=triangle]
        plot coordinates {
        (250,329.2)
        (500,659.6)
        (1000,1324.4)
        (2000,2645.6)
      };
      \addlegendentry{${D4_i}$}
    \end{axis}
  \end{tikzpicture}
  \caption{Comparison of the average number of moves for the C4/D4 algorithms on random trees of various orders.}
  \label{fig:c4}
\end{figure*}

\begin{figure*}
  \begin{tikzpicture}
    \begin{axis}[
      xlabel=Order $n$,
      ylabel=\% Difference,
      legend pos=outer north east]
      \addplot[smooth,mark=*,color=dark-gray]
        plot coordinates {
        (250,33.50)
        (500,32.7)
        (1000,32.22)
        (2000,32.19)
      };
      \addlegendentry{${{C4_\beta}_0}$ vs ${C1}$}

      \addplot[smooth,color=dark-gray,mark=x]
        plot coordinates {
        (250,62.02)
        (500,61.9)
        (1000,62.33)
        (2000,61.51)
      };
      \addlegendentry{${C4_i}$ vs ${C1}$}

      \addplot[smooth,color=dark-gray,mark=square]
        plot coordinates {
        (250,27.46)
        (500,27.5)
        (1000,27.32)
        (2000,27.39)
      };
      \addlegendentry{${{D4_\beta}_0}$ vs ${D1}$}

      \addplot[smooth,color=dark-gray,mark=triangle]
        plot coordinates {
        (250,51.63)
        (500,51.8)
        (1000,51.94)
        (2000,52.04)
      };
      \addlegendentry{${D4_i}$ vs ${D1}$}
    \end{axis}
  \end{tikzpicture}
  \caption{Comparison of the \% differences for the average number of moves of the C4/D4 algorithms with C1/D1 on random trees of various orders.}
  \label{fig:c4Comparison}
\end{figure*}

Our simulation results indicate that the central scheduler algorithms usually find larger independent sets than the distributed scheduler algorithms, see Table \ref{tab:tSetSize}, and generally require fewer moves, see Table \ref{tab:tCompSteps}, to reach a stable state. The second result was expected, as distributed algorithms often include wait states. However, we were not expecting the first result. Of the algorithms studied, Algorithm C2, and its distributed version D2, are faster than the previously published algorithms C1 and D1. In terms of set size, the following inequality was observed for finding large maximal independent sets:

\begin{equation}
C4, D4 > C3, D3 > C2, D2 > C1 = C5, D1 = D5 
\end{equation}

and the following inequality for small maximal independent sets:

\begin{equation}
C4, D4 < C3, D3 < C2, D2 < C1 = C5, D1 = D5 
\end{equation}

Algorithm C4, and the distributed D4, consistently produced the largest and smallest maximal independent sets. This is also consistent with findings in \cite{vt16}. Algorithm C5, and its distributed version D5, produced sets of the same size as C1 and the D1 but made far more moves.


\end{section}


\begin{section}{Conclusions}\label{Conclusion_Section}

With the introduction of distributed sensors in environments such as the smart warehouse, the need for algorithms that can be run on these sensors in order to create topological layouts, such as maximal independent sets, will only increase. This paper has presented a variety of algorithms and approaches for determining fast methods for finding a MIS set as well as methods for determining the smallest/largest MIS set.

The complexity of Algorithm D4, as well as a fixed bound for Algorithm C4, are still open problems and of interest to the authors.

\end{section}





\bibliographystyle{elsarticle-num}







\end{document}